\documentclass[12pt]{article}
\usepackage{latexsym}
\usepackage{graphicx}
\usepackage{caption2}
\usepackage{psfrag}
\usepackage{amsmath}
\newcommand{\be}{\begin{equation}}
\newcommand{\ee}{\end{equation}}
\newcommand{\bea}{\begin{eqnarray}}
\newcommand{\eea}{\end{eqnarray}}

\newcommand{\lesssim}{ {\
\lower-1.2pt\vbox{\hbox{\rlap{$<$}\lower5pt\vbox{\hbox{$\sim$}}}}\ } }
\newcommand{\gtrsim}{ {\
\lower-1.2pt\vbox{\hbox{\rlap{$>$}\lower5pt\vbox{\hbox{$\sim$}}}}\ } }
\input epsf

\begin{document}

\begin{titlepage}

\begin{flushright}
%\today
\end{flushright}
\vspace*{1.5cm}
\begin{center}
{\Large \bf Large-$N_c$ QCD and Pade Approximant Theory}\\[3.0cm]

{\bf Santiago Peris }

% and {\bf YYY }\\[1cm]

 Grup de F{\'\i}sica Te{\`o}rica and IFAE\\ Universitat Aut{\`o}noma de Barcelona, 08193 Barcelona, Spain.\\[0.5cm]

\end{center}

\vspace*{1.0cm}

\begin{abstract}

In the large-$N_c$ limit of QCD, the expansion of the vacuum polarization at low
energies determines the whole function at any arbitrarily large (but finite) energy.
This result is an immediate consequence of the Theory of Pade Approximants to
Stieltjes functions.

\end{abstract}

\end{titlepage}

In the limit of a infinite number of colors, QCD simplifies and becomes a theory of
stable noninteracting mesons\cite{Hooft}. The number of these mesons  is infinite
because Green's functions have to reproduce the logarithmic dependence on energy
present in the parton model. In fact, in spite of all the attempts made, the
solution of QCD in the large $N_c$ limit has not yet been found. The large-$N_c$
limit of QCD is still too difficult to be solved exactly. It seems to be
interesting, therefore, to develop a systematic approximation to QCD in this limit.
In this paper, we would like to point out that Pade Approximants are such an
approximation, at least in certain cases.

Given a function $f(z)$, of complex variable $z$, with power series expansion about
the origin given by
\begin{equation}\label{pade2}
    f(z)\approx \sum_{n=0}^{N+M} f_{n} z^n\ ,
\end{equation}
we shall define the Pade Approximant, $R^{N}_{M}(z)$, as the rational function
\begin{equation}\label{pade1}
    R^{N}_{M}(z)= \frac{\sum_{n=0}^{N} a_n z^n}{\sum_{n=0}^{M} b_n z^n}\
\end{equation}
satisfying the condition that its expansion about $z=0$ matches $N+M+1$ terms in the
expansion of the function $f(z)$ in Eq. (\ref{pade2}). Notice that, without loss of
generality, one can always choose the first coefficient $b_0=1$.

In general, the convergence properties of Pade Approximants are much more difficult
than those of, e.g., power series and it is not guaranteed that they will converge
to the function $f(z)$ in the combined limit $N,M\rightarrow \infty$. This of course
calls into question the reliability of the approximation and probably explains why
they are not such a popular tool in Quantum Field Theory as compared to power
series.

However, there is a very important exception to this general situation. If the
function $f(z)$ is a Stieltjes function whose expansion about the origin has a
finite radius of convergence $R>0$ , i.e.
\begin{equation}\label{pade3}
    f(z)=\int_{0}^{1/R} \frac{d\nu(\tau)}{1+\tau z} \ ,
\end{equation}
with $\nu(\tau)$ a bounded, nondecreasing function\footnote{Notice that $\nu(t)$ is
not even required to be continuous.} with finite real-valued moments given by
\begin{equation}\label{pade4}
    f_n=(-1)^n \int_{0}^{1/R} \tau^n\ d\nu (\tau)\ ,
\end{equation}
then there is a theorem in the Theory of Pade Approximants which assures convergence
for $N=M+J$, $J\geq -1$, as $M\rightarrow \infty$ in Eqs.
(\ref{pade2}-\ref{pade1})\cite{Baker}.

As we shall now see, the vacuum polarization function is an example of a Stieltjes
function of the momentum variable. To simplify matters, let us take the isospin
limit ($m_u=m_d$) and define $\Pi_V(Q^2)$, with $q^2=-Q^2$, as
\begin{equation}\label{vector}
    (q_{\mu} q_{\nu} - q^2 g_{\mu\nu})\Pi_V(Q^2)=\int d^4x\ e^{i qx} \langle
    T\{\overline{u}\gamma_{\mu}d(x)\overline{d}\gamma_{\nu}u(0)\}\rangle\ .
\end{equation}
As it is very well known, the function $\Pi_V(Q^2)$ satisfies a once-subtracted
dispersion relation given by
\begin{equation}\label{disp}
    \Pi_V(Q^2)=\Pi_V(0)- \frac{Q^2}{\pi}
    \int_{M_{\rho}^2}^{\infty}\frac{dt}{t(t+Q^2)}\  \mathrm{Im}\Pi_V(t) \ ,
\end{equation}
where $M_{\rho}^2$ is the mass of the lowest vector state, the rho meson. In the
large-$N_c$ limit, which we have already taken in Eq. (\ref{disp}), the resonance
spectrum is made of infinitely narrow one-particle states. This is the reason why
the lower limit is cutoff at $M_{\rho}^2$ in Eq. (\ref{disp}). Of course, we are
assuming here that there is a mass gap in large-$N_c$ QCD, i.e. $M_{\rho}^2\neq 0$.

For any finite $N_c$, two-particle states also contribute, which gives the spectral
function a support all the way down to $t=4 m_{\pi}^2$. However, the physics between
the two scales  $m_{\pi}$ and $M_{\rho}$ is already well understood in terms of
Chiral Perturbation Theory\cite{GL}. It is the physics in the intermediate scales
between $M_{\rho}$ and the onset of the QCD perturbative continuum  at a few GeV
which remains difficult. In Chiral Perturbation Theory, all the physics at and above
the first resonance mass is encoded in a set of low-energy constants which, in
principle, depend on the dynamics of QCD, although the precise connection is
unknown. As the papers in Ref. \cite{Ecker,Ecker2} show, it is for this connection
that the large-$N_c$ expansion may become most useful. Therefore, in the following
we shall only consider the $N_c\rightarrow \infty$ limit.

The basic observation is actually extremely simple. Making the change of variable
$t=1/\tau$ in Eq. (\ref{disp}), one can rewrite this combination as
\begin{equation}\label{thirtyseven}
   \Phi(Q^2)\equiv \frac{\Pi_V(0)-\Pi_V(Q^2)}{Q^2}=
   \int_{0}^{\frac{1}{M_{\rho}^2}} \frac{d\nu(\tau)}{1+\tau Q^2}\ ,
\end{equation}
with $d\nu\left(\tau \right)= \frac{1}{\pi} \mathrm{Im}\Pi_V(\tau^{-1}) d\tau$ with
$\mathrm{Im}\Pi_V(\tau^{-1})\geq 0$, i.e. one can express the function $\Phi(Q^2)$
precisely in the form (\ref{pade3}), with $R=M_{\rho}^2$. The integral
representation (\ref{thirtyseven}) assures that the combination $\Phi(Q^2)$ is a
Stieltjes function of the complex variable $Q^2$ and allows one to make connection
with Pade Approximant Theory.

The expansion of $\Phi(Q^2)$ about the origin is related to the corresponding Chiral
Perturbation Theory expansion\cite{GL} and can be written like
\begin{equation}\label{chi}
    \Phi(Q^2)=\sum_{n=0}^{\infty} \mathrm{L}_{n} (Q^2)^n\ .
\end{equation}
In the chiral limit ($m_u,m_d\rightarrow 0$) the coefficients $\mathrm{L}_{n}$ are
given by the appropriate low energy constants which  accompany the operators with
chiral dimension $2n+6$ in the chiral Lagrangian. These coefficients are given in
terms of the spectral function by the following integral
\begin{equation}\label{spec}
    \mathrm{L}_n= (-1)^n  \int_{0}^{\frac{1}{M_{\rho}^2}}d\tau \tau^n\
    \frac{1}{\pi}\mathrm{Im}\Pi_V(\tau^{-1})\ .
\end{equation}
Due to the upper limit in the integral representation (\ref{thirtyseven}), clearly
the power series expansion (\ref{chi}) is convergent only for $|Q^2|< M_{\rho}^2$.
We would like to emphasize that the underlying reason for the existence of the
simple power series in $Q^2$  given by (\ref{chi}) below $M_{\rho}^2$, i.e. with no
chiral logarithms, is the cutoff at $M_{\rho}$ at the lower end of the spectrum in
(\ref{disp}) which is in turn a consequence of the $N_c\rightarrow \infty$ limit. As
is well known, chiral logarithms are subleading effects at large $N_c$
\cite{GL,Ecker,Ecker2}.

The $N_c\rightarrow\infty$ limit of the vector spectral function becomes the highly
singular distribution, i.e.
\begin{equation}\label{spectral}
    \frac{1}{\pi}\mathrm{Im}\Pi_V(t)= \sum_{n=1}^{\infty} F_n^2 M_n^2 \delta (t-M_n^2)\ ,
\end{equation}
with the index $n$ labeling the increasingly heavier sequence of vector resonances,
i.e. $M_n=M_{\rho},M_{\rho'},M_{\rho''},...$ as $n=1,2,3,...$ Consequently, the
combination (\ref{thirtyseven}) is a meromorphic function and can be written as the
infinite sum
\begin{equation}\label{combo}
    \Phi(Q^2)= \sum_{n=1}^{\infty}\frac{F_n^2}{Q^2+M_n^2}\ ,
\end{equation}
in terms of the resonance masses, $M_n$, and decay constants, $F_n$.

On the other hand, the Pade  $R^{M+J}_{M}(Q^2)$ can always be decomposed using
partial fractions as
\begin{equation}\label{partfrac}
    R^{M+J}_{M}(Q^2) = P^{[J/M]}(Q^2) + \sum_{r=1}^{M} \frac{a_{r}}{Q^2+ \mu^2_r}\ ,
\end{equation}
where $P^{[J/M]}(Q^2)$ is a polynomial of degree $J$ when $J\geq 0$, and it vanishes
if $J<0$. The above mentioned theorem\cite{Baker} states that the Pade
$R^{M+J}_{M}(Q^2)$ in Eq. (\ref{pade1}) will converge to $\Phi(Q^2)$ as
$M\rightarrow \infty$, for any $J\geq -1$. In fact, the convergence takes place for
any finite value of $Q^2$ in the whole $Q^2$ complex plane except on the negative
real axis  $-\infty <Q^2<-M^2_{\rho}$, where the poles of $\Phi(Q^2)$, i.e. the
resonance masses, are located. If the Stieltjes function under consideration is
multivalued (as is the case of the vacuum polarization function due to the presence
of logarithms), the limit $M\rightarrow \infty$ of the Pade sequence selects the
branch which is real for $Q^2$ real and positive\cite{Bender}. This is precisely the
physical definition of $\Phi(Q^2)$. All the poles of $R_{M}^{M+J}(Q^2)$ lie on the
negative real axis with $-\infty <Q^2<-M^2_{\rho}$. Furthermore, they are simple
poles and with positive residues\cite{Baker2}.

It is quite remarkable that out of the low-energy chiral expansion given in
(\ref{chi}) one may reconstruct the full function $\Phi(Q^2)$. Notice, in
particular, that this function contains logarithms of the variable $Q^2$ at
euclidean high energies because the anomalous dimensions of the relevant operators
in the operator product expansion do not vanish, in general, in the large-$N_c$
limit. The above convergence theorem means that all this complexity is reproduced by
the rational approximant (\ref{partfrac}) as $M\rightarrow \infty$. Pades offer a
proper mathematical way to approximate certain large-$N_c$ QCD functions, such as
the vacuum polarization $\Phi(Q^2)$ in (\ref{combo}), by means of a rational
function, i.e. with a finite number of poles. We think that these results may give a
new perspective to some phenomenological approximations used in the past, starting
with the celebrated vector meson dominance\cite{Ecker}.

Some results about the rate of convergence are known. For instance, in the case of
the $R^{M-1}_{M}$ Pade one has that\cite{Baker3}
\begin{equation}\label{fortyfive}
    \left|\Phi(Q^2)- R^{M-1}_{M}(Q^2)
    \right| \leq K\ \left| \frac{\sqrt{M_{\rho}^2+Q^2}-\sqrt{M_{\rho}^2}}
    {\sqrt{M_{\rho}^2+Q^2}+\sqrt{M_{\rho}^2}}
    \right|^{2M}\ ,
\end{equation}
where $K$ is an unspecified constant. The bound is valid for any complex $Q^2$,
provided it is not on the cut $-\infty<Q^2<-M_{\rho}^2$. Notice that the numerator
in the bound (\ref{fortyfive}) is smaller than the denominator and, therefore, it
goes to zero as $M\rightarrow \infty$.

A word of caution is required, however. The fact that the approximant
(\ref{partfrac}) converges to the full function $\Phi(Q^2)$ for any finite value of
$Q^2$ does not mean that one can \emph{expand} this approximant in powers of $1/Q^2$
at high energies. On the contrary, any such expansion cannot be exact as it is
unable to reproduce the $\log Q^2$ behavior due to the nonvanishing anomalous
dimensions. In other words, the limits $Q^2\rightarrow \infty$ and $M\rightarrow
\infty$ do not commute, as it is clear from Eq. (\ref{fortyfive}).

In this respect let us mention that a completely different subject is the use of
the, so-called, two-point Pade Approximants\cite{Baker5}. These are also rational
functions similar to (\ref{pade1}), but for which the construction of the two
polynomials is done by matching simultaneously both expansions of the function
$f(z)$ around $z=0$ and $z=\infty$. In this case, if the function $f(z)$ has a power
series expansion in $1/z$ at infinity, these powers will of course be trivially
reproduced by construction. However, as we have already emphasized, Green's
functions in QCD have logarithms in the large $Q^2$ expansion. Strictly speaking,
this presents an obstruction to the construction of these two-point Pades. In
certain cases where the Green's function considered vanishes to all orders of
perturbation theory\footnote{Such as, e.g., the left-right vacuum polarization
function $\Pi_{LR}$ in the chiral limit.}, these logarithms enter only as a
correction to the leading $1/Q^2$ power and they are screened by at least a power of
the coupling constant $\alpha_s$. It becomes then justifiable to neglect these logs
in a first approximation\cite{Ecker2}. In this case, however, the Pades no longer
approximate the full function and their convergence properties are not known.

In the case of the vacuum polarization function (\ref{combo}), the fact that
$R_{M}^{M+J}(Q^2)$ converges when  $M\rightarrow \infty$ means that the polynomial $
P^{[J/M]}(Q^2) $ in Eq. (\ref{partfrac}) tends to zero in this limit. Consequently,
out of all the possible Pade approximants, the $R^{M-1}_{M}(Q^2)$ Pade is
particularly interesting because it allows a direct comparison of its poles and
residues with the resonance masses and decay constants appearing in the function
$\Phi(Q^2)$ in Eq. (\ref{combo}).

Restricting ourselves to this Pade $R^{M-1}_{M}$ in Eq. (\ref{partfrac}) (i.e. with
$P^{[J/M]}(Q^2)=0$), one can easily find relations among resonance masses and
low-energy constants. For instance, one can obtain a first-order approximation to
the $\rho$ mass as the position of the only pole of the lowest-order Pade
$R^{0}_{1}(Q^2)$. This turns out to be located at
\begin{equation}\label{pole1}
    \mu^2_1=- \frac{\mathrm{L_0}}{\mathrm{L_1}}\ .
\end{equation}
This prediction can be refined by looking at the he next-to-leading Pade,
$R^{1}_{2}(Q^2)$. Its two poles at  $\mu^2_1$ and $ \mu^2_2$ yield approximate
values for the  $\rho$ and  $\rho'$ masses which verify the relations
\begin{equation}\label{pole2}
    \mu_1^2 + \mu_2^2=-\ \frac{\mathrm{L}_1 \mathrm{L_2} - \mathrm{L}_0
    \mathrm{L}_3}{\mathrm{L}_2^2-\mathrm{L}_1 \mathrm{L}_3}\qquad ,\qquad
    \frac{1}{\mu_1^2}+ \frac{1}{\mu_2^2}=-\
    \frac{\mathrm{L}_1 \mathrm{L}_2 -
    \mathrm{L}_0 \mathrm{L}_3}{\mathrm{L}_1^2- \mathrm{L}_0 \mathrm{L}_2}\ .
\end{equation}
The above prediction for the first two resonances is not equally good for both of
them, however. Because the Pade approximation worsens as one moves away from the
origin, one should expect that the prediction for a resonance mass is better the
lower the resonance. This means better predictions for the $\rho$ than for the
$\rho'$, and the latter better than for the $\rho''$, etc... for a given Pade.

Should more terms from the chiral expansion be known, we could in principle go on
refining the above predictions in a systematic way. Regretfully, the $\mathrm{L}_n$
are presently not known. Nevertheless, in principle these terms can be determined in
the future by analyzing scattering processes of pions and kaons in Chiral
Perturbation Theory\cite{GL}, or by other methods such as, e.g., the lattice. When
this will be done, it will be interesting to check the above relations.

In the mean time we shall content ourselves with a simple illustration of some of
these results by means of a toy model. To this end, let us consider the following
function defined by the sum\cite{Shifman,Phily}

%%%%%%%%%%%%%%%%%%%%%%%%%%%%%%%%%%%%%%%%%%%%%%%%%%%%%%%%%%%%%%%%%%%%%%%%%
%%%%%%%%%  EXAMPLE OF FIGURE  %%%%%%%%%%%%%%%%%%%%%%%%%%%%%%%%%%%%%%%%%%%
\begin{figure}
\renewcommand{\captionfont}{\small \it}
\renewcommand{\captionlabelfont}{\small \it}
\centering
\includegraphics[width=3in]{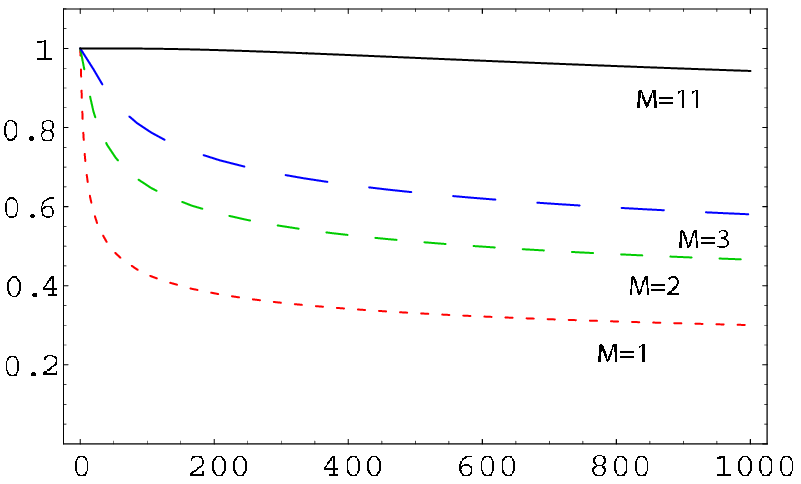}
\hspace{0.in}
\includegraphics[width=3in]{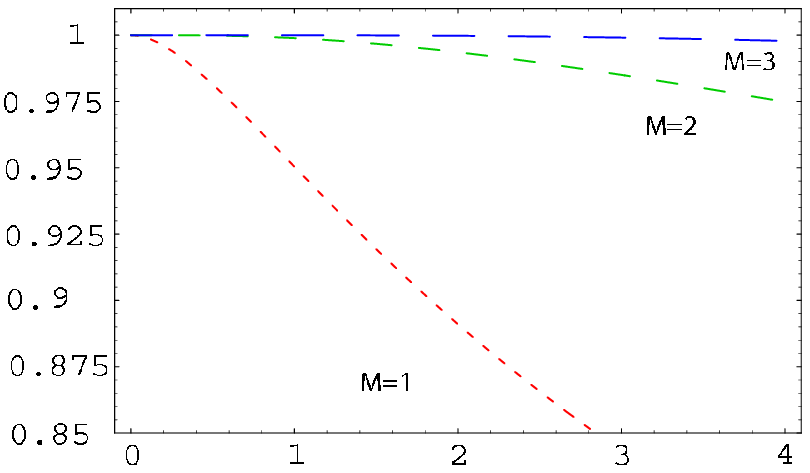}
\caption{(Left) Plot of the ratio $\frac{R^{M-1}_{M}(Q^2)}{\Phi_{Toy}(Q^2)}$ as a
function of $Q^2$, for $M=1,2,3$ and $M=11$. Notice the large values of $Q^2$
considered. (Right) Close-up of the same curves with $M=1,2,3$ in the low-$Q^2$
region.}\label{fig:resonanceloop}
\end{figure}
%%%%%%%%%%%%%%%%%%%%%%%%%%%%%%%%%%%%%%%%%%%%%%%%%%%%%%%%%%%%%%%%%%%%%%%%%

\begin{equation}\label{toy}
    \Phi_{Toy}(Q^2)=\sum_{n=1}^{\infty} \frac{1}{n (Q^2+n)}\ ,
\end{equation}
as a very simple toy model for the once-subtracted vacuum polarization
(\ref{thirtyseven}) in the large-Nc limit. Obviously its spectrum consists of masses
located at $M^2_n=1,2,3,...$, and one may use the mass of the lowest resonance as
the unit of energy. An exact result for $\Phi_{Toy}(Q^2)$ can be readily obtained as
\begin{equation}\label{exact}
    \Phi_{Toy}(Q^2)=\frac{1}{Q^2}\Big\{\psi(Q^2+1)+ \gamma\Big\}
\end{equation}
where the function $\psi(Q^2+1)$ is the Digamma function, defined as
\begin{equation}\label{def}
\psi(z)=\frac{d}{dz} \log \Gamma (z)\ ,
\end{equation}
with $\Gamma (z)$ the Euler Gamma function and $\gamma\simeq 0.577216$ being the
Euler-Mascheroni constant. The function $\Phi_{Toy}(Q^2)$ has the ``chiral''
expansion
\begin{equation}\label{exactexpand}
    \Phi_{Toy}(Q^2)\approx \sum_{p=0}^{\infty} (-1)^p \zeta(p+2)\ Q^{2p}\ ,
\end{equation}
where $\zeta(p+2)=\sum_{n=1}^{\infty} n^{-p-2}$ is the  Riemann's $\zeta$ function.
On the other hand, the large $Q^2$ expansion
 is given by ($|\arg Q^2|<\pi$)

\begin{equation}\label{exactexpand2}
     \Phi_{Toy}(Q^2)\approx \frac{\gamma + \log Q^2}{Q^2}+ \frac{1}{2 Q^4}-
     \sum_{p=1}^{\infty} \frac{B_{2p}}{2p\ Q^{4p+2}}\ ,
\end{equation}
where $B_{2p}=(-1)^{p+1} \frac{2 (2p)!}{(2\pi)^{2p}}\ \zeta(2p)$ are the Bernoulli
numbers for $p=1,2,3,...$.

 While the chiral expansion has a finite radius of convergence $|Q^2|< 1$,
the large $Q^2$ expansion has zero radius of convergence (i.e. it is asymptotic) due
to the factorial growth of the Bernoulli numbers. Notice also the presence of the
$\log Q^2$ behavior in (\ref{exactexpand2}). Even so, Fig. 1 shows how the Pade
Approximants  $R^{M-1}_{M}$ are able to reproduce the full function at any finite
value of $Q^2$. Notice that no step function simulating a perturbative ``continuum''
is needed \cite{Phily}. The fact that they approach the true function from below is
also  well-known \cite{Bender}.

As to the prediction for the resonance masses, Table 1 shows the position of the
poles of the given Pades and compares them to the true values for the masses. As
anticipated before, the prediction for the masses gets worse the heavier the
resonance, for a given Pade. The prediction for a low resonance, however, is
certainly reasonable even for a low-order Pade and becomes much better as the order
of the Pade increases.

\begin{table}
\centering
\begin{tabular}{|c|c|c|c|c|c|c|}
  \hline
   & $R^{0}_{1}$ & $R^{1}_{2}$ & $R^{2}_{3}$ & $R^{3}_{4}$ & $R^{4}_{5}$ & $R^{5}_{6}$\\
  \hline
  $M_1^2=1$ & 1.3684 &1.0228  &1.0010  &1.0000 & 1. & 1.\\
  $ M_2^2=2$ &  &4.3711  &2.2967  &2.0409 & 2.0038 & 2.0002\\
   $M_3^2=3$ &  &  & 10.1201 & 4.1755&  3.2698 & 3.0526\\
   $M_4^2=4$ &  &  &  & 19.5578 & 6.9160 & 4.8688\\
  \hline
\end{tabular}
\caption{\emph{Squared masses of the resonances obtained as poles of the
corresponding Pade. For comparison, the true squared masses are also shown on the
leftmost column.}}\label{label}
\end{table}

In conclusion, Pade Approximants, when properly used, may turn out to be very useful
tools for studying large-$N_c$ QCD. We have exemplified this in the case of the
vacuum polarization because, having a positive spectral function, it happens to be a
Stieltjes function, to which a known convergence theorem applies. But this case is
not unique. Similar results will also apply to other correlators with positive
spectral functions. Apart from this, it would certainly be most interesting to find
out under which conditions Pade Approximants converge in more general situations
when the spectral function is not positive definite.

The so-called two-point Pades also seem to be very promising, as they are able to
connect the Operator Product and Chiral expansions\cite{Ecker2}. However, to our
knowledge, the consequences of their convergence properties for a QCD Green's
function have not yet been analyzed. It is clearly tempting to speculate about the
possibility that they may be linked to theories with extra dimensions\cite{Erlich}.

\vspace{1 cm}

\textbf{Acknowledgements}

I am grateful to E. de Rafael for encouragement and numerous discussions, to F.J.
Yndurain for correspondence and to M. Golterman for a critical reading of the
manuscript. Finally, I would also like to thank P. Gonzalez-Vera for very useful
conversations about Pades. This work has been supported by
CICYT-FEDER-FPA2005-02211, SGR2005-00916 and by TMR, EC-Contract No.
HPRN-CT-2002-00311 (EURIDICE).


\begin{thebibliography}{99}

\bibitem{Hooft}
  G.~'t Hooft,
  Nucl.\ Phys.\ B {\bf 72} (1974) 461,
  %%CITATION = NUPHA,B72,461;%%
 %``A Two-Dimensional Model For Mesons,''
  Nucl.\ Phys.\ B {\bf 75} (1974) 461;
  %%CITATION = NUPHA,B75,461;%%
E.~Witten,
  %``Baryons In The 1/N Expansion,''
  Nucl.\ Phys.\ B {\bf 160} (1979) 57.
  %%CITATION = NUPHA,B160,57;%%

\bibitem{Baker}
G.A. Baker and P. Graves-Morris, \textit{Pade Approximants}, Encyclopedia of
Mathematics and its Applications, Cambridge Univ. Press 1996. Section 5.4, Theorem
5.4.2.

\bibitem{GL}
See, e.g., J.~Gasser and H.~Leutwyler,
  %``Chiral Perturbation Theory: Expansions In The Mass Of The Strange Quark,''
  Nucl.\ Phys.\ B {\bf 250} (1985) 465.
  %%CITATION = NUPHA,B250,465;%%




\bibitem{Bender}
C. Bender and S. Orszag, \textit{Advanced Mathematical Methods for Scientists and
Engineers I: asymptotic methods and perturbation theory}, Springer 1999, section 8.6


\bibitem{Baker2}
G.A. Baker and P. Graves-Morris, Ref. \cite{Baker}. Section 5.2, Theorem 5.2.1.


\bibitem{Ecker}
See, e.g.,  G.~Ecker, J.~Gasser, A.~Pich and E.~de Rafael,
  %``The Role Of Resonances In Chiral Perturbation Theory,''
  Nucl.\ Phys.\ B {\bf 321} (1989) 311;
  %%CITATION = NUPHA,B321,311;%%
 J.~F.~Donoghue, C.~Ramirez and G.~Valencia,
  %``The Spectrum Of QCD And Chiral Lagrangians Of The Strong And Weak
  %Interactions,''
  Phys.\ Rev.\ D {\bf 39} (1989) 1947;
  %%CITATION = PHRVA,D39,1947;%%
 S.~Peris, M.~Perrottet and E.~de Rafael,
  %``Matching long and short distances in large-N(c) {QCD},''
  JHEP {\bf 9805} (1998) 011
  [arXiv:hep-ph/9805442].
  %%CITATION = HEP-PH 9805442;%%



\bibitem{Baker3}
G.A. Baker, \textit{Essentials of Pade Approximants}, Academic Press 1975. Chapter
16, Theorem 16.2; see also, G.A. Baker and P. Graves-Morris, Ref.\cite{Baker},
Section 5.4, Theorem 5.4.4.


\bibitem{Baker5}
G.A. Baker and P. Graves-Morris, Ref.\cite{Baker}, Chapter 7.

\bibitem{Ecker2}
G.~Ecker, J.~Gasser, H.~Leutwyler, A.~Pich and E.~de Rafael,
  %``Chiral Lagrangians For Massive Spin 1 Fields,''
  Phys.\ Lett.\ B {\bf 223} (1989) 425;
  %%CITATION = PHLTA,B223,425;%%
  M.~Knecht and E.~de Rafael,
  %``Patterns of spontaneous chiral symmetry breaking in the large N(c)  limit
  %of QCD-like theories,''
  Phys.\ Lett.\ B {\bf 424} (1998) 335
  [arXiv:hep-ph/9712457];
  %%CITATION = HEP-PH 9712457;%%
  M.~Knecht, S.~Peris and E.~de Rafael,
  %``The electroweak pi+ pi0 mass difference and weak matrix elements in  the
  %1/N(c) expansion,''
  Phys.\ Lett.\ B {\bf 443} (1998) 255
  [arXiv:hep-ph/9809594];
  %%CITATION = HEP-PH 9809594;%%
 E.~de Rafael,
  %``Analytic approaches to kaon physics,''
  Nucl.\ Phys.\ Proc.\ Suppl.\  {\bf 119} (2003) 71
  [arXiv:hep-ph/0210317];
  %%CITATION = HEP-PH 0210317;%%
  S.~Peris,
  %``Electroweak matrix elements at large N(c): Matching quarks to mesons,''
  arXiv:hep-ph/0204181;
  %%CITATION = HEP-PH 0204181;%%
 A.~Pich,
  %``Present status of chiral perturbation theory,''
  Int.\ J.\ Mod.\ Phys.\ A {\bf 20} (2005) 1613
  [arXiv:hep-ph/0410322].
  %%CITATION = HEP-PH 0410322;%%

\bibitem{Shifman}
B.~Blok, M.~A.~Shifman and D.~X.~Zhang,
  %``An illustrative example of how quark-hadron duality might work,''
  Phys.\ Rev.\ D {\bf 57}, 2691 (1998)
  [Erratum-ibid.\ D {\bf 59}, 019901 (1999)]
  [arXiv:hep-ph/9709333].
  %%CITATION = HEP-PH 9709333;%%


\bibitem{Phily}
See, e.g., M.~Golterman, S.~Peris, B.~Phily and E.~de Rafael,
  %``Testing an approximation to large-N(c) QCD with a toy model,''
  JHEP {\bf 0201} (2002) 024
  [arXiv:hep-ph/0112042].
  %%CITATION = HEP-PH 0112042;%%

\bibitem{Erlich}
 J.~Erlich, G.~D.~Kribs and I.~Low,
  %``Emerging holography,''
  arXiv:hep-th/0602110;
  %%CITATION = HEP-TH 0602110;%%
J.~Hirn and V.~Sanz,
  %``Interpolating between low and high energy QCD via a 5D Yang-Mills  model,''
  JHEP {\bf 0512} (2005) 030
  [arXiv:hep-ph/0507049];
  %%CITATION = HEP-PH 0507049;%%
L.~Da Rold and A.~Pomarol,
  %``Chiral symmetry breaking from five dimensional spaces,''
  Nucl.\ Phys.\ B {\bf 721} (2005) 79
  [arXiv:hep-ph/0501218].
  %%CITATION = HEP-PH 0501218;%%






\end{thebibliography}
\end{document}